\documentclass[aps,jmp,reprint,prl,showpacs]{revtex4-1}
\usepackage{graphicx}
\usepackage{amsmath}
\usepackage{amssymb}
\usepackage{dcolumn}
\usepackage{multirow}
\usepackage{hhline}
\usepackage{blindtext}
\usepackage{braket}
\usepackage{siunitx}
\usepackage[acronym]{glossaries}
\usepackage[utf8]{inputenc}
\usepackage{color}
\newacronym{2deg}{2DEG}{two-dimensional electron gas}
\newacronym{qpc}{QPC}{quantum point contact}
\newacronym[longplural={quantum dots}]{qd}{QD}{quantum dot}
\newacronym[longplural={double quantum dots}]{DQD}{DQD}{double quantum dot}
\newacronym{awg}{AWG}{arbitrary waveform generator}
\newacronym{dos}{DOS}{density of states}
\newcommand{\up}{\ensuremath{\uparrow}}
\newcommand{\down}{\ensuremath{\downarrow}}

\newacronym{SOI}{SOI}{spin--orbit interaction}

\begin{document}
\preprint{ETH Zurich}

\title{Anisotropy and suppression of spin--orbit interaction in a GaAs double quantum dot} 

\author{A. Hofmann}
\email[]{andrea.hofmann@phys.ethz.ch}
\author{V. F. Maisi}
\author{T. Krähenmann}
\author{C. Reichl}
\author{W. Wegscheider}
\author{K. Ensslin}
\author{T. Ihn}
\affiliation{Laboratory for Solid State Physics, ETH Zurich}

\date{\today}
\textsl{}
\begin{abstract}
The spin-flip tunneling rates are measured in GaAs-based double quantum dots by time-resolved charge detection. Such processes occur in the Pauli spin blockade regime with two electrons occupying the double quantum dot. Ways are presented for tuning the spin-flip tunneling rate, which on the one hand gives access to measuring the Rashba and Dresselhaus spin--orbit coefficents. On the other hand they make it possible to turn on and off the effect of \acrlong{SOI} with a high on/off-ratio. The tuning is accomplished by choosing the alignment of the tunneling direction with respect to the crystallographic axes, as well as by choosing the orientation of the external magnetic field with respect to the spin--orbit magnetic field. Spin-lifetimes of $\SI{10}{\second}$ are achieved at a tunnel rate close to $\SI{1}{\kilo\hertz}$.
\end{abstract}

\maketitle 

\Gls{SOI} couples the orbital motion of electrons to the spin via electric fields. Electrons moving in a crystal experience spin--orbit coupling originating from electric fields with bulk (Dresselhaus) and structure (Rashba) inversion asymmetries. Theoretical work predicts the \gls{SOI} to bring about interesting physical phenomena, such as the quantum spin Hall effect \cite{kane_quantum_2005,murakami_dissipationless_2003,sinova_universal_2004} and, in conjuction with superconductivity, Majorana states in nanowires \cite{sau_generic_2010}. Large \gls{SOI} can be used for driving spin qubits \cite{nowack_coherent_2007}, and vanishing spin--orbit magnetic fields are required for observing a persistent spin-helix \cite{koralek_emergence_2009,bernevig_exact_2006,sasaki_direct_2014}. Considering the varying influence of \gls{SOI} in different systems, measuring its anisotropy and finding ways to tune its strength are essential for taking advantage of it.

\begin{figure}[h!]
	\includegraphics[width = \linewidth]{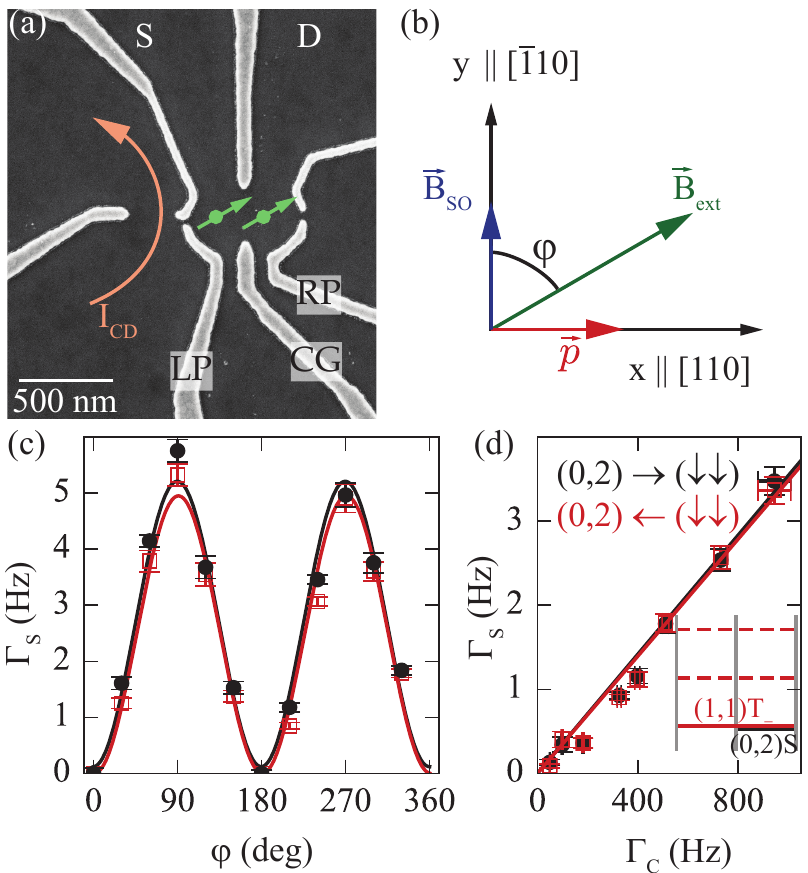}
	\caption{\label{fig:Setup} (a) A scanning electron micrograph of \gls{DQD}1, top-gates shown in light grey, zero volts are applied between source (S) and drain (D). The current $I_\textrm{CD}$ distinguishes the charge states (1,1) and (0,2) of the two electron \gls{DQD}. (b) Crystallographic orientation of the device in (a) including the direction $\vec{p}$ of tunneling, the spin--orbit magnetic field $\vec{B}_\mathrm{SO}$ and the angle $\varphi$ between the externally applied in-plane field $\vec{B}_\mathrm{ext}$ and the spin--orbit field. (c) Spin-flip tunneling rate versus the angle between the spin--orbit magnetic field $\vec{B}_\mathrm{SO}$ and the external field $\vec{B}_\mathrm{ext}$. Points with error bars are experimental data, solid lines are fits to theory. The fitted values at $\varphi=\SI{0}{\degree},\SI{180}{\degree}$ amount to $\SI[separate-uncertainty=true]{0.20(17)}{\hertz}$ (black line) and $\SI[separate-uncertainty=true]{0.07(17)}{\hertz}$ (red line). (d) Spin-flip tunneling rate measured as a function of the spin-conserving tunneling rate $\Gamma_\mathrm{C}$. Inset: resonant alignment of the $(1,1)\mathrm{T}_-$ and the $(0,2)$S \gls{DQD} states.}
\end{figure}


Despite the relevance of SOI for various experimental systems, only few experiments exist testing the effect of its anisotropy on spin relaxation in quantum dots which are candidates for qubits.~\cite{loss_quantum_1998,imamoglu_quantum_1999}. It has been studied in single quantum dots \cite{amasha_electrical_2008,scarlino_spin-relaxation_2014,konemann_spin-orbit_2005,golovach_phonon-induced_2004,golovach_spin_2008,khaetskii_spin_2000,khaetskii_spin-flip_2001,falko_anisotropy_2005,stano_orbital_2006} with undefined direction of electron momentum and large energy gap between the different spin states \cite{scarlino_spin-relaxation_2014}. In coupled quantum dots, theoretical studies predict anisotropic singlet--triplet splitting \cite{burkard_cancellation_2002,stepanenko_spin-orbit_2003,stano_orbital_2006,golovach_spin_2008,stepanenko_singlet-triplet_2012,kavokin_anisotropic_2001,danon_spin-flip_2013}, and experimental evidence has been obtained recently in highly coupled dots but with ambiguity about the crystallographic direction of the main DQD axis defining the electron momentum \cite{nichol_quenching_2015}. In this paper, we probe the \gls{SOI} in GaAs by measuring spin-flip tunneling of individual electrons between energetically resonant (1,1) and (0,2) charge states of double quantum dots [see Fig.~1(a)] and with well-defined direction of electron tunneling. A magnetic field is applied in the plane of the underlying two-dimensional electron gas. In particular, we experimentally explore the two-fold anisotropy of the spin--orbit magnetic field
\begin{align}
\label{eq:somag}
\vec{B}_\textrm{SO} = 2 \left( (\alpha-\beta)p_y,(\alpha+\beta)p_x,0\right) /(g \mu_B)
\end{align}
with respect to the crystallographic direction $\vec{p}$ of electron motion, as well as with respect to the spin quantization axis given by the external magnetic field. In Eq.~\eqref{eq:somag}, $\alpha$ and $\beta$ are the Rashba- and Dresselhaus coefficients, $\mu_\mathrm{B}$ is Bohr's magneton, and $g$ is the effective conduction band $g$-factor of GaAs. Our results yield a thorough experimental verification of theoretically anti\-ci\-pated \cite{stepanenko_singlet-triplet_2012,golovach_phonon-induced_2004} \gls{SOI} effects, and allow for a strong suppression of spin-flip tunneling for electrons moving along the $[\bar{1}10]$ crystal direction as well as for external magnetic fields parallel to the internal spin--orbit field.

The relevance of \gls{SOI} for electrons tunneling between two quantum dots in a GaAs/Al$_{.3}$Ga$_{.7}$As heterostructure is determined by the three factors illustrated in Fig.~\ref{fig:Setup}(b). First, there is the orientation of the crystal lattice defining the principal axes $[110]$ ($x$-direction) and $[\bar{1}10]$ ($y$-direction) in the (001)-plane of the electron gas. Second, there is the direction $\vec{p}$ of tunneling between the quantum dots, which determines the orientation of the spin--orbit magnetic field $\vec{B}_\mathrm{SO}$ in the same plane [Eq.~\eqref{eq:somag}], as experienced by the tunneling electron. In our experiment we have chosen the tunneling direction to be along $[110]$ in device \gls{DQD}1 [shown in Fig.~\ref{fig:Setup}(a)] and along $[\bar{1}10]$ in a nominally identical device \gls{DQD}2 (not shown). Third, there is the quantization axis of the spins in the initial and final quantum dot states, which is chosen in our experiment by applying a magnetic field $\vec{B}_\mathrm{ext}$ in the $(001)$-plane at varying angles $\varphi$ with respect to $[\bar{1}10]$ ($[110]$) for \gls{DQD}1 (\gls{DQD}2), i.e. with respect to the expected direction of $\vec{B}_\textrm{SO}$. In an intuitive classical picture \cite{stano_spin-orbit_2005} the spin--orbit magnetic field $\vec{B}_\mathrm{SO}$ causes the spin to precess around the spin-quantization axis during the tunneling process. The precession angle, and therefore the spin-flip tunneling rate is maximum for $\vec{B}_\mathrm{SO}\perp\vec{B}_\mathrm{ext}$, minimum for $\vec{B}_\mathrm{SO}\parallel\vec{B}_\mathrm{ext}$, and varies sinusoidally in-between. This oscillation provides the relevant handle for tuning the spin--orbit coupling strength in situ, as we will show in the following. By using a DQD for our study, the direction of motion of the tunnelling electron is well-defined and we can measure the influence of the SOI on possible qubit operations. In contrast to Ref.~\cite{nichol_quenching_2015}, the main DQD axis and therefore the direction of motion is known to point along $[110]$ for DQD1 and along $[\bar{1}10]$ for DQD2. The two devices show qualitatively different spin-flip rates. In particular, we find suppressed SOI for DQD2.

We apply $B_\textrm{ext}=\SI{1.5}{\tesla}$, such that the (1,1)T$_-$ spin-triplet state with spin-configuration $\ket{\down\down}$ is the lowest energy state with one electron in each dot as shown in the inset of Fig.~\ref{fig:Setup}(d). The large Zeeman splitting of $\SI{38}{\micro\electronvolt}$, much larger than $k_\mathrm{B}T=\SI{4.3}{\micro\electronvolt}$ at the electronic temperature $T_\textrm{el}=\SI{50}{\milli\kelvin}$, suppresses excitations to energetically higher $(1,1)$ spin-triplet states, and thereby freezes the spin-orientation within the individual QDs. We tune the spin-singlet state $(0,2)$S with two electrons in the right dot into resonance with $(1,1)\mathrm{T}_-$ [see Fig.~\ref{fig:Setup}(d), inset], while suppressing tunneling to source and drain to a rate $\leq\SI{0.1}{\hertz}$. Due to the different spin alignments of these states, each tunneling event between them requires a spin-flip. Using time-resolved detection of single-electron tunneling we measure the spin-flip tunneling rate $\Gamma_\mathrm{SO}$ in \gls{DQD}1 for the two individual resonant transitions $(1,1)\mathrm{T}_-\leftrightarrow(0,2)\mathrm{S}$. Varying the angle between $\vec{B}_\mathrm{ext}$ and $\vec{B}_\mathrm{SO}$, we plot the two rates in Fig.~\ref{fig:Setup}(c) in red and black. Apparently, the spin-flip tunneling is completely suppressed for $\vec{B}_\mathrm{ext}\parallel\vec{B}_\mathrm{SO}$ and significantly enhanced for $\vec{B}_\mathrm{ext}\perp\vec{B}_\mathrm{SO}$. The suppression at parallel alignment is so strong that we observe no transitions on a time scale up to $\SI{10}{\second}$ even though the tunnel coupling between the resonant states would allow for spin-conserving tunneling at a rate of about \SI{1}{\kilo\hertz}, as explained below. This is the most significant data reported in this paper.

By means of a charge-state dependent feedback \cite{hofmann_measuring_2016} (an explanation of this follows later in this paper) we additionally measure the spin-conserving tunneling rate $\Gamma_\mathrm{C}=\SI{1}{\kilo\hertz}$. Figure~\ref{fig:Setup}(d) shows that the measured spin-flip tunneling rate $\Gamma_\mathrm{SO}$ is in this regime proportional to the spin-conserving tunneling rate $\Gamma_\mathrm{C}$. We vary this quantity by changing the voltage applied to the central gate CG [see Fig.~\ref{fig:Setup}(a)]. It has indeed been theoretically proposed \cite{golovach_phonon-induced_2004,danon_spin-flip_2013} that the spin-flip tunneling rate is given by
\begin{align}
	\label{eq:soratetheory}
	\Gamma_\textrm{SO}&=\frac{d^2}{2 l_\textrm{SO}^2} \Gamma_\textrm{C} \sin^2(\varphi),
\end{align}
where $d$ is the inter-dot distance and $l_\textrm{SO}$ the spin--orbit length \cite{golovach_phonon-induced_2004}. The spin--orbit length $l_\textrm{SO}=\hbar/m^*(\alpha \pm \beta)$ is a measure for the strength of the \gls{SOI}. It is different for the two devices studied here: the plus sign is valid for tunneling along $[110]$ (\gls{DQD}1), and the minus sign for tunneling along $[\bar{1}10]$ (\gls{DQD}2). The numerical values of the Rashba ($\alpha)$ and Dresselhaus ($\beta$) spin--orbit coefficients are unknown a-priori, and $m^* = 0.067 m_e$ is the electron mass in GaAs. Our experimental results do indeed confirm the prediction \eqref{eq:soratetheory} as we show in Figs.~\ref{fig:Setup}(c) and (d) by indicating fits to the theoretical prediction with red and black solid lines.

\begin{figure}
\includegraphics[width = \linewidth]{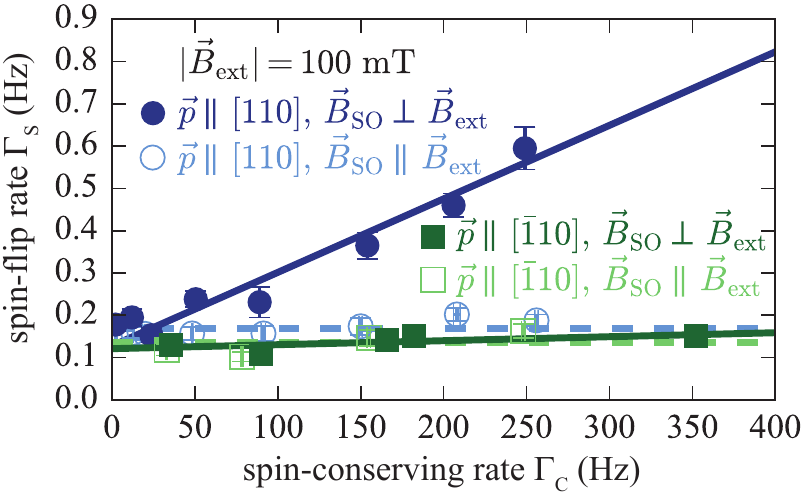}
\caption{\label{fig:Rate} The spin-flip rate as a function of the spin-conserving rate at an externally applied magnetic field of $100$~mT. The blue circles represent data taken from \gls{DQD}1 ($\vec{p}\parallel\,[110]$) and green squares are used for \gls{DQD}2 ($\vec{p}\parallel\,[\bar{1}10]$). Filled symbols are used for perpendicular alignment of the spin quantization axis and the spin--orbit field, and open symbols for the parallel alignment. The suppression of the \gls{SOI} is visible in the vanishing dependence of the spin-flip rate on the tunnel coupling.}
\end{figure}

We now demonstrate that we can gradually increase the importance of dot-internal spin relaxation \cite{golovach_phonon-induced_2004,stano_orbital_2006,hanson_spins_2007,scarlino_spin-relaxation_2014} by reducing the strength of the external magnetic field $B_\mathrm{ext}$ to $\SI{100}{\milli\tesla}$. This reduces the Zeeman energy to $\SI{2.5}{\micro\electronvolt}$, which is now smaller than the thermal energy. Due to equipartitioning \cite{reichl_modern_2016}, all four $(1,1)$ spin states, [$\ket{\up\up},\ket{\up\down},\ket{\down\up},\ket{\down\down}$] are occupied with similar probability. Spin-flip tunneling processes in this regime can be distinguished by their proportionality to $\Gamma_\mathrm{C}$ from dot-internal spin relaxation with subsequent tunneling, which is independent of $\Gamma_\mathrm{C}$ \cite{maisi_spin-orbit_2016}. The blue filled circles in Fig.~\ref{fig:Rate} show the total spin-flip rate $\Gamma_\mathrm{S}$ with $\vec{B}_\mathrm{ext}\parallel [110]$ for \gls{DQD}1 as $\Gamma_\mathrm{C}$ is changed. The linear dependence differs from that in Fig.~\ref{fig:Setup}(d) by a vertical offset indicating a finite spin-flip rate within the individual quantum dots. In turn, aligning $\vec{B}_\mathrm{ext}$ along $[\bar{1}10]$ in the same device leads to the blue open circles, which show essentially no dependence on $\Gamma_\mathrm{C}$ in agreement with $\varphi=0$ in Eq.~\eqref{eq:soratetheory}, but have the same offset as the data of the $\varphi=90^\circ$ case. It is evident from the data that we can write the total measured spin-flip rate as
\begin{align}
\label{eq:intext}
\Gamma_\mathrm{S} = \Gamma_\mathrm{int} + \Gamma_\mathrm{SO}(\varphi),
\end{align}
where $\Gamma_\mathrm{int}$ is the dot-internal spin-flip rate. As we observe this rate to be similar in the two \gls{DQD} devices, we deduce that the hyperfine spin environment and the spin--orbit coupling strengths within a single dot [hence within the charge state (1,1)] are similar in all the dots. In contrast to Ref.~\cite{scarlino_spin-relaxation_2014}, we do not observe a large anisotropy in the dot-internal spin-flip rate. With the small Zeeman splitting in our experiment, dot-internal mixing of spin-states is primarily caused by hyperfine interaction \cite{fujita_signatures_2016}, and thermal activation to all states is possible. The spin-flip processes in this setting differs from the spin-relaxation process from the excited spin-state to the ground-state with an energy gap of $7\,k_\textrm{B}T$ studied in Ref.~\cite{scarlino_spin-relaxation_2014}. In our experiments, we are interested in the spin-orbit mediated rate $\Gamma_\mathrm{SO}$, which is measured by the contribution linear in $\Gamma_\mathrm{C}$ to the spin-flip rate $\Gamma_\mathrm{S}$, as shown in Fig.~\ref{fig:Rate}.

Comparing these measurements taken on \gls{DQD}1 with those taken on \gls{DQD}2, where the direction $\vec{p}$ of tunneling is along $[\bar{1}10]$, gives the filled and open green squares in Fig.~\ref{fig:Rate}. Both orientations of $\vec{B}_\mathrm{ext}$, namely $\varphi=0$ and $\varphi=90^\circ$ give datapoints without a pronounced $\Gamma_\mathrm{C}$-dependence. This behavior can be attributed to the different values of $l_\mathrm{SO}$ in the two devices. From the vanishing \gls{SOI} in \gls{DQD}2, we conclude that $\alpha - \beta \approx 0$~\cite{studer_gate-controlled_2009,koralek_emergence_2009,dettwiler_stretchable_2017,schliemann_nonballistic_2003,bernevig_exact_2006} in our electron gas, which is device specific in general~\cite{koralek_emergence_2009} and accidental in our case. Comparing the slopes of the two devices measured at $\varphi=\ang{90}$, we determine an upper bound for the relative difference of the Rashba and Dresselhaus spin--orbit contribution and find that the \gls{SOI} for electrons moving along $[\bar{1}10]$ is suppressed by a factor $\left|(\alpha-\beta)/(\alpha+\beta)\right|^2 < \SI{6e-2}{}$ compared to the $[110]$ direction. Thus, the alignment of the \gls{DQD} main axis with respect to the crystal axes allows for the selection of the spin--orbit field experienced by the tunneling electrons. With a separation $d \approx \SI{300}{\nano\meter}$ of the two quantum dots, we estimate $\alpha \approx \beta \approx \SI{1.5e3}{\meter\per\second}$, which agrees well with spin--orbit lengths of a few $\SI{}{\micro m}$ found in the literature \cite{nowack_coherent_2007,nichol_quenching_2015,maisi_spin-orbit_2016}.

\begin{figure}
	\includegraphics[width = \linewidth]{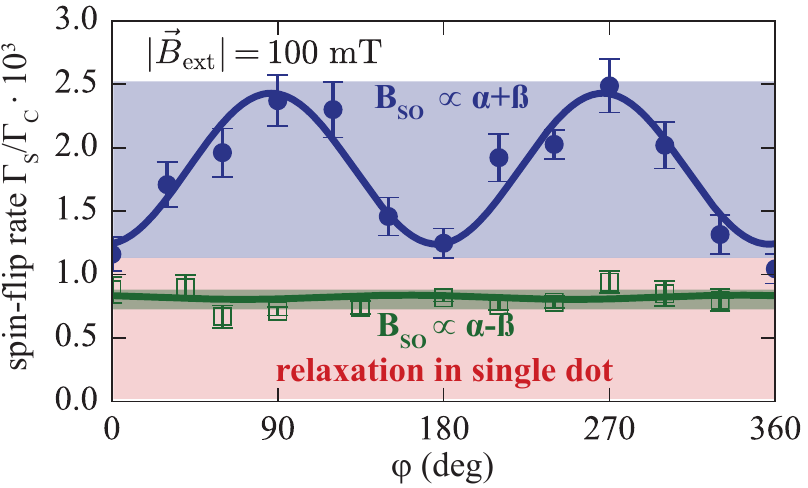}
	\caption{\label{fig:Angle} The spin-flip rate normalized to the spin-conserving rate as a function of the in-plane angle $\varphi$ between the external field (spin quantization axis) and the spin--orbit field (perturbation). The sinus\-oidal dependence found in \gls{DQD}1 (filled blue circles) reflects the anisotropy of the perturbation induced by the \gls{SOI}. The \gls{SOI} vanishes for \gls{DQD}2 (open green squares), leading to an almost isotropic spin-flip rate. The isotropic background (red) is determined by the spin-flip processes within the $(1,1)$ charge state in the individual \gls{DQD} devices.}
\end{figure}

Figure~\ref{fig:Angle} shows the full angle-dependence of $\Gamma_\mathrm{S}$ for both devices with $\Gamma_\mathrm{C}=\SI{160}{\hertz}$. In agreement with Eq.~\eqref{eq:intext}, \gls{DQD}1 (filled blue circles) shows the strong oscillatory contribution to $\Gamma_\mathrm{S}$ originating from $\Gamma_\mathrm{SO}(\varphi)$ given in Eq.~\eqref{eq:soratetheory}, but vertically offset by $\Gamma_\mathrm{int}$. In contrast, the open green squares of \gls{DQD}2 show only a very weak angle-dependence, in agreement with the weak spin--orbit interaction effect experienced by an electron tunneling in $[\bar{1}10]$ direction (c.f. Fig.~\ref{fig:Rate}). The solid lines are sinus\-oidal fits to Eqs.~\eqref{eq:intext} and \eqref{eq:soratetheory} with $\Gamma_\textrm{int}$, $d^2/2l_\textrm{SO}^2$ and a phase offset $\varphi_0$ as fitting parameters. The phase offset accounts for an uncertainty in the alignment of the sample with respect to the magnetic field and is $\varphi_0<\ang{5}$ in \gls{DQD}1. In this measurement, the rate $\Gamma_\textrm{C}=\SI{160}{\hertz}$ was found to be independent of the angle $\varphi$. Summing up our results demonstrated in Figs.~\ref{fig:Rate} and \ref{fig:Angle}, we used Eq.~\eqref{eq:intext} to distinguish spin-flip tunnelling from dot-internal relaxation processes. We showed that the spin-orbit interaction is suppressed for an electron tunneling along $[\bar{1}10]$. For tunneling along $[110]$, spin-flip tunnelling is reduced for the field directions $\varphi=\SI{0}{\degree}$ and $\SI{180}{\degree}$.

\begin{figure}
	\includegraphics[width = \linewidth]{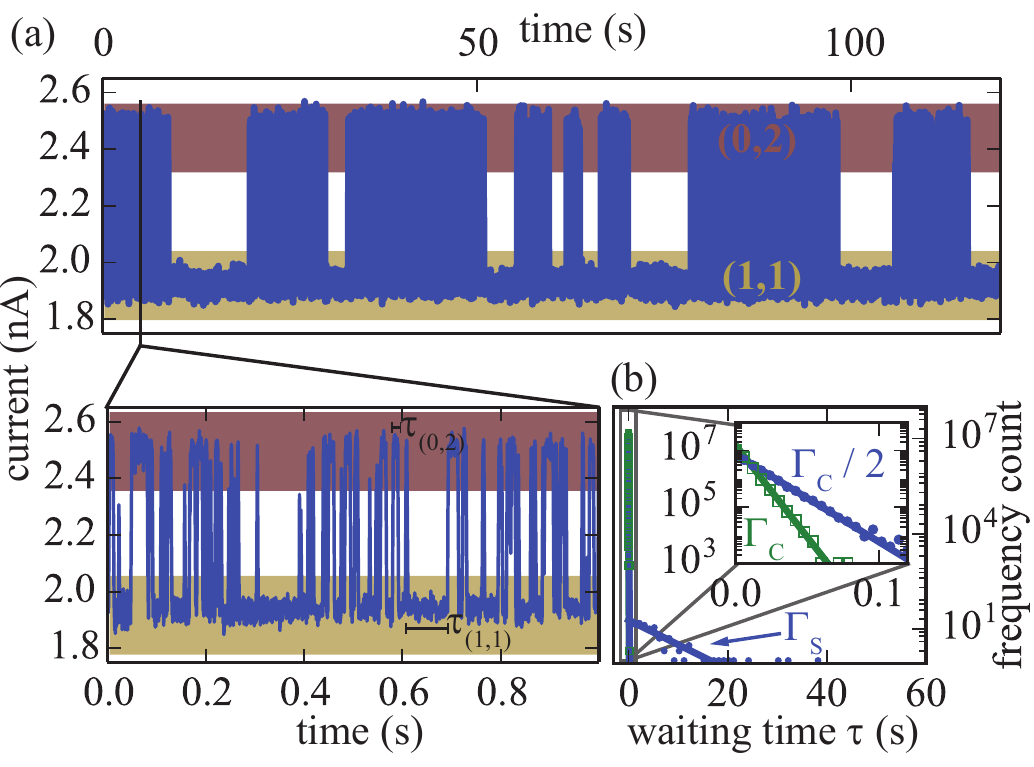}
	\caption{\label{fig:Trace} Charge detection. (a) The charge detector current $I_\textrm{CD}$ changes between two values, depending on the charge state of the two-electron \gls{DQD}. The bunching of tunneling events is visible in the long time trace, and the zoom shows the fast tunneling events when the electron spins are antiparallel. The waiting time distribution is plotted in (b), where green and blue symbols are used for the $(0,2)$ and $(1,1)$ state, respectively. The slow time-scale (spin-flip rate) is visible in the main figure and the fast time-scale (spin-conserving rates) in the zoom-in.}
\end{figure}

We now present the technical details of our experiment and of the measurements of $\Gamma_\mathrm{S}$ and $\Gamma_\mathrm{C}$. Our devices are formed in a GaAs/AlGaAs heterostructure containing a two-dimensional electron gas $\SI{90}{\nano\metre}$ below the surface. The electron density and mobility are $n=\SI{2e11}{cm^{-2}}$ and $\mu=\SI[per-mode=symbol]{2e6}{{\centi\metre}^2\per{\volt\second}}$, respectively, as measured at $\SI{1.3}{\kelvin}$. Negative voltages applied to the metallic top gates [light-grey fingers in Fig.~\ref{fig:Setup}(a)] deplete the two-dimensional electron gas below the gates and confine the electrons (green circles with arrows) in \gls{DQD}s \cite{hanson_spins_2007}.

We measure the charge state of the \gls{DQD} by means of the current $I_\textrm{CD}$ through a quantum point contact charge detector coupled capacitively to the \gls{DQD}-device [see Fig.~\ref{fig:Setup}(a)] \cite{vandersypen_real-time_2004,schleser_time-resolved_2004}. The time-trace of $I_\textrm{CD}$ presented in Fig.~\ref{fig:Trace}(a) was taken with $B_\mathrm{ext}=100\,\mathrm{mT}$, i.e. larger than the Overhauser field \cite{jouravlev_electron_2006,koppens_control_2005}. It shows transitions between the $(1,1)$ and $(2,0)$ states. In the topmost panel, we observe that occasionally the tunneling becomes halted for several seconds as a result of Pauli spin blockade \cite{ono_current_2002,johnson_singlet-triplet_2005,fujita_signatures_2016,fujita_single_2015,maisi_spin-orbit_2016} at small magnetic fields: if the spins of the two electrons in the $(1,1)$ state are parallel [spin states $\ket{\down\down}$ or $\ket{\up\up}$], a spin-flip is required for a transition to the $(2,0)$ state, where the spins of the electrons are anti-parallel \cite{hanson_spins_2007}. On the other hand, fast tunneling events occur between the $(0,2)$ and the unpolarized $(1,1)$ states [spin states $\ket{\up\down}, \ket{\down\up}$], where no spin-flip is required. The distribution of waiting times $\tau_{(1,1)}$ in the $(1,1)$ state [see the zoom of Fig.~\ref{fig:Trace}(a) for the definition of this quantity] is presented in Fig.~\ref{fig:Trace}(b) with filled blue circles and shows two characteristic time scales, see also Refs.~\cite{maisi_spin-orbit_2016,fujita_signatures_2016}. We associate the fast one with the spin conserving tunneling rate, $\Gamma_\textrm{C}$, set by the tunnel-coupling between the two quantum dots. The slow time-scale is associated with transitions requiring a spin-flip and is a measure for the spin-flip rate $\Gamma_\textrm{S}$. 

Finally we briefly explain the operation of the charge-dependent feedback mechanism used to determine $\Gamma_\mathrm{C}$ in the data shown in Fig.~\ref{fig:Setup}. When both electrons are detected in the same dot, the $(0,2)$ level is shifted into resonance with the unpolarized $(1,1)$ states. The measured waiting time for tunneling into one of these states yields the spin-conserving rate $\Gamma_\textrm{C}$. The rate $\Gamma_\mathrm{C}$ weakly depends on the in-plane angle $\varphi$ with a sinusoidal amplitude smaller than $\SI{14}{\percent}$. Possible origins are the confinement of the wave--functions perpendicular to the field direction, and spin-flip tunnelling combined with a relaxation into the $(1,1)T_-$ state.

In conclusion, we have demonstrated experimentally the two-fold anisotropy of the \acrlong{SOI}. We showed that the strength of the spin--orbit field depends on the direction of electron tunneling and, in our device, vanishes for the electron tunneling along $[\bar{1}10]$. We were able to extract values for the spin--orbit coefficients $\alpha$ and $\beta$ from our observations. Along two well-defined directions of electron tunneling, we measured the anisotropy of the spin--orbit interaction brought about by the relative alignment between spin quantization axis and the spin--orbit field and observed the theoretically expected sinusoidal dependence. Our measurements at magnetic fields, where the Zeeman splitting exceeds temperature, demonstrate the suppression of spin relaxation processes within single dots. The high tunability of the spin-flip tunneling rate and the absence of the incoherent processes found in this configuration are promising for coherent spin operations in \gls{DQD}s. Beyond that our measurement technique lends itself for similar studies in different material systems, where few or no studies of the anisotropy of spin--orbit interaction are available.

\begin{acknowledgments}
We acknowledge contributions by Simon Parolo. We thank Mattias Beck, Peter Stano, Gian Salis and Patrick Altmann for stimulating discussions and Daniel Loss for valuable feedback on the manuscript. This work was supported by the Swiss National Science Foundation (SNF) through the National Center of Competence in Research Quantum Science and Technology (NCCR QSIT) and by Eidgenössische Technische Hochschule (ETH) Zürich.
\end{acknowledgments}
\bibliography{bibliography}

\begin{thebibliography}{41}%
\makeatletter
\providecommand \@ifxundefined [1]{%
 \@ifx{#1\undefined}
}%
\providecommand \@ifnum [1]{%
 \ifnum #1\expandafter \@firstoftwo
 \else \expandafter \@secondoftwo
 \fi
}%
\providecommand \@ifx [1]{%
 \ifx #1\expandafter \@firstoftwo
 \else \expandafter \@secondoftwo
 \fi
}%
\providecommand \natexlab [1]{#1}%
\providecommand \enquote  [1]{``#1''}%
\providecommand \bibnamefont  [1]{#1}%
\providecommand \bibfnamefont [1]{#1}%
\providecommand \citenamefont [1]{#1}%
\providecommand \href@noop [0]{\@secondoftwo}%
\providecommand \href [0]{\begingroup \@sanitize@url \@href}%
\providecommand \@href[1]{\@@startlink{#1}\@@href}%
\providecommand \@@href[1]{\endgroup#1\@@endlink}%
\providecommand \@sanitize@url [0]{\catcode `\\12\catcode `\$12\catcode
  `\&12\catcode `\#12\catcode `\^12\catcode `\_12\catcode `\%12\relax}%
\providecommand \@@startlink[1]{}%
\providecommand \@@endlink[0]{}%
\providecommand \url  [0]{\begingroup\@sanitize@url \@url }%
\providecommand \@url [1]{\endgroup\@href {#1}{\urlprefix }}%
\providecommand \urlprefix  [0]{URL }%
\providecommand \Eprint [0]{\href }%
\providecommand \doibase [0]{http://dx.doi.org/}%
\providecommand \selectlanguage [0]{\@gobble}%
\providecommand \bibinfo  [0]{\@secondoftwo}%
\providecommand \bibfield  [0]{\@secondoftwo}%
\providecommand \translation [1]{[#1]}%
\providecommand \BibitemOpen [0]{}%
\providecommand \bibitemStop [0]{}%
\providecommand \bibitemNoStop [0]{.\EOS\space}%
\providecommand \EOS [0]{\spacefactor3000\relax}%
\providecommand \BibitemShut  [1]{\csname bibitem#1\endcsname}%
\let\auto@bib@innerbib\@empty
\bibitem [{\citenamefont {Kane}\ and\ \citenamefont
  {Mele}(2005)}]{kane_quantum_2005}%
  \BibitemOpen
  \bibfield  {author} {\bibinfo {author} {\bibfnamefont {C.~L.}\ \bibnamefont
  {Kane}}\ and\ \bibinfo {author} {\bibfnamefont {E.~J.}\ \bibnamefont
  {Mele}},\ }\href {\doibase 10.1103/PhysRevLett.95.226801} {\bibfield
  {journal} {\bibinfo  {journal} {Phys. Rev. Lett.}\ }\textbf {\bibinfo
  {volume} {95}},\ \bibinfo {pages} {226801} (\bibinfo {year}
  {2005})}\BibitemShut {NoStop}%
\bibitem [{\citenamefont {Murakami}\ \emph {et~al.}(2003)\citenamefont
  {Murakami}, \citenamefont {Nagaosa},\ and\ \citenamefont
  {Zhang}}]{murakami_dissipationless_2003}%
  \BibitemOpen
  \bibfield  {author} {\bibinfo {author} {\bibfnamefont {S.}~\bibnamefont
  {Murakami}}, \bibinfo {author} {\bibfnamefont {N.}~\bibnamefont {Nagaosa}}, \
  and\ \bibinfo {author} {\bibfnamefont {S.-C.}\ \bibnamefont {Zhang}},\ }\href
  {\doibase 10.1126/science.1087128} {\bibfield  {journal} {\bibinfo  {journal}
  {Science}\ }\textbf {\bibinfo {volume} {301}},\ \bibinfo {pages} {1348}
  (\bibinfo {year} {2003})}\BibitemShut {NoStop}%
\bibitem [{\citenamefont {Sinova}\ \emph {et~al.}(2004)\citenamefont {Sinova},
  \citenamefont {Culcer}, \citenamefont {Niu}, \citenamefont {Sinitsyn},
  \citenamefont {Jungwirth},\ and\ \citenamefont
  {MacDonald}}]{sinova_universal_2004}%
  \BibitemOpen
  \bibfield  {author} {\bibinfo {author} {\bibfnamefont {J.}~\bibnamefont
  {Sinova}}, \bibinfo {author} {\bibfnamefont {D.}~\bibnamefont {Culcer}},
  \bibinfo {author} {\bibfnamefont {Q.}~\bibnamefont {Niu}}, \bibinfo {author}
  {\bibfnamefont {N.~A.}\ \bibnamefont {Sinitsyn}}, \bibinfo {author}
  {\bibfnamefont {T.}~\bibnamefont {Jungwirth}}, \ and\ \bibinfo {author}
  {\bibfnamefont {A.~H.}\ \bibnamefont {MacDonald}},\ }\href {\doibase
  10.1103/PhysRevLett.92.126603} {\bibfield  {journal} {\bibinfo  {journal}
  {Phys. Rev. Lett.}\ }\textbf {\bibinfo {volume} {92}},\ \bibinfo {pages}
  {126603} (\bibinfo {year} {2004})}\BibitemShut {NoStop}%
\bibitem [{\citenamefont {Sau}\ \emph {et~al.}(2010)\citenamefont {Sau},
  \citenamefont {Lutchyn}, \citenamefont {Tewari},\ and\ \citenamefont
  {Das~Sarma}}]{sau_generic_2010}%
  \BibitemOpen
  \bibfield  {author} {\bibinfo {author} {\bibfnamefont {J.~D.}\ \bibnamefont
  {Sau}}, \bibinfo {author} {\bibfnamefont {R.~M.}\ \bibnamefont {Lutchyn}},
  \bibinfo {author} {\bibfnamefont {S.}~\bibnamefont {Tewari}}, \ and\ \bibinfo
  {author} {\bibfnamefont {S.}~\bibnamefont {Das~Sarma}},\ }\href {\doibase
  10.1103/PhysRevLett.104.040502} {\bibfield  {journal} {\bibinfo  {journal}
  {Phys. Rev. Lett.}\ }\textbf {\bibinfo {volume} {104}},\ \bibinfo {pages}
  {040502} (\bibinfo {year} {2010})}\BibitemShut {NoStop}%
\bibitem [{\citenamefont {Nowack}\ \emph {et~al.}(2007)\citenamefont {Nowack},
  \citenamefont {Koppens}, \citenamefont {Nazarov},\ and\ \citenamefont
  {Vandersypen}}]{nowack_coherent_2007}%
  \BibitemOpen
  \bibfield  {author} {\bibinfo {author} {\bibfnamefont {K.~C.}\ \bibnamefont
  {Nowack}}, \bibinfo {author} {\bibfnamefont {F.~H.~L.}\ \bibnamefont
  {Koppens}}, \bibinfo {author} {\bibfnamefont {Y.~V.}\ \bibnamefont
  {Nazarov}}, \ and\ \bibinfo {author} {\bibfnamefont {L.~M.~K.}\ \bibnamefont
  {Vandersypen}},\ }\href {\doibase 10.1126/science.1148092} {\bibfield
  {journal} {\bibinfo  {journal} {Science}\ }\textbf {\bibinfo {volume}
  {318}},\ \bibinfo {pages} {1430} (\bibinfo {year} {2007})}\BibitemShut
  {NoStop}%
\bibitem [{\citenamefont {Koralek}\ \emph {et~al.}(2009)\citenamefont
  {Koralek}, \citenamefont {Weber}, \citenamefont {Orenstein}, \citenamefont
  {Bernevig}, \citenamefont {Zhang}, \citenamefont {Mack},\ and\ \citenamefont
  {Awschalom}}]{koralek_emergence_2009}%
  \BibitemOpen
  \bibfield  {author} {\bibinfo {author} {\bibfnamefont {J.~D.}\ \bibnamefont
  {Koralek}}, \bibinfo {author} {\bibfnamefont {C.~P.}\ \bibnamefont {Weber}},
  \bibinfo {author} {\bibfnamefont {J.}~\bibnamefont {Orenstein}}, \bibinfo
  {author} {\bibfnamefont {B.~A.}\ \bibnamefont {Bernevig}}, \bibinfo {author}
  {\bibfnamefont {S.-C.}\ \bibnamefont {Zhang}}, \bibinfo {author}
  {\bibfnamefont {S.}~\bibnamefont {Mack}}, \ and\ \bibinfo {author}
  {\bibfnamefont {D.~D.}\ \bibnamefont {Awschalom}},\ }\href {\doibase
  10.1038/nature07871} {\bibfield  {journal} {\bibinfo  {journal} {Nature}\
  }\textbf {\bibinfo {volume} {458}},\ \bibinfo {pages} {610} (\bibinfo {year}
  {2009})}\BibitemShut {NoStop}%
\bibitem [{\citenamefont {Bernevig}\ \emph {et~al.}(2006)\citenamefont
  {Bernevig}, \citenamefont {Orenstein},\ and\ \citenamefont
  {Zhang}}]{bernevig_exact_2006}%
  \BibitemOpen
  \bibfield  {author} {\bibinfo {author} {\bibfnamefont {B.~A.}\ \bibnamefont
  {Bernevig}}, \bibinfo {author} {\bibfnamefont {J.}~\bibnamefont {Orenstein}},
  \ and\ \bibinfo {author} {\bibfnamefont {S.-C.}\ \bibnamefont {Zhang}},\
  }\href {\doibase 10.1103/PhysRevLett.97.236601} {\bibfield  {journal}
  {\bibinfo  {journal} {Phys. Rev. Lett.}\ }\textbf {\bibinfo {volume} {97}},\
  \bibinfo {pages} {236601} (\bibinfo {year} {2006})}\BibitemShut {NoStop}%
\bibitem [{\citenamefont {Sasaki}\ \emph {et~al.}(2014)\citenamefont {Sasaki},
  \citenamefont {Nonaka}, \citenamefont {Kunihashi}, \citenamefont {Kohda},
  \citenamefont {Bauernfeind}, \citenamefont {Dollinger}, \citenamefont
  {Richter},\ and\ \citenamefont {Nitta}}]{sasaki_direct_2014}%
  \BibitemOpen
  \bibfield  {author} {\bibinfo {author} {\bibfnamefont {A.}~\bibnamefont
  {Sasaki}}, \bibinfo {author} {\bibfnamefont {S.}~\bibnamefont {Nonaka}},
  \bibinfo {author} {\bibfnamefont {Y.}~\bibnamefont {Kunihashi}}, \bibinfo
  {author} {\bibfnamefont {M.}~\bibnamefont {Kohda}}, \bibinfo {author}
  {\bibfnamefont {T.}~\bibnamefont {Bauernfeind}}, \bibinfo {author}
  {\bibfnamefont {T.}~\bibnamefont {Dollinger}}, \bibinfo {author}
  {\bibfnamefont {K.}~\bibnamefont {Richter}}, \ and\ \bibinfo {author}
  {\bibfnamefont {J.}~\bibnamefont {Nitta}},\ }\href {\doibase
  10.1038/nnano.2014.128} {\bibfield  {journal} {\bibinfo  {journal} {Nat
  Nano}\ }\textbf {\bibinfo {volume} {9}},\ \bibinfo {pages} {703} (\bibinfo
  {year} {2014})}\BibitemShut {NoStop}%
\bibitem [{\citenamefont {Loss}\ and\ \citenamefont
  {DiVincenzo}(1998)}]{loss_quantum_1998}%
  \BibitemOpen
  \bibfield  {author} {\bibinfo {author} {\bibfnamefont {D.}~\bibnamefont
  {Loss}}\ and\ \bibinfo {author} {\bibfnamefont {D.~P.}\ \bibnamefont
  {DiVincenzo}},\ }\href {\doibase 10.1103/PhysRevA.57.120} {\bibfield
  {journal} {\bibinfo  {journal} {Phys. Rev. A}\ }\textbf {\bibinfo {volume}
  {57}},\ \bibinfo {pages} {120} (\bibinfo {year} {1998})}\BibitemShut
  {NoStop}%
\bibitem [{\citenamefont {Imamoglu}\ \emph {et~al.}(1999)\citenamefont
  {Imamoglu}, \citenamefont {Awschalom}, \citenamefont {Burkard}, \citenamefont
  {DiVincenzo}, \citenamefont {Loss}, \citenamefont {Sherwin},\ and\
  \citenamefont {Small}}]{imamoglu_quantum_1999}%
  \BibitemOpen
  \bibfield  {author} {\bibinfo {author} {\bibfnamefont {A.}~\bibnamefont
  {Imamoglu}}, \bibinfo {author} {\bibfnamefont {D.~D.}\ \bibnamefont
  {Awschalom}}, \bibinfo {author} {\bibfnamefont {G.}~\bibnamefont {Burkard}},
  \bibinfo {author} {\bibfnamefont {D.~P.}\ \bibnamefont {DiVincenzo}},
  \bibinfo {author} {\bibfnamefont {D.}~\bibnamefont {Loss}}, \bibinfo {author}
  {\bibfnamefont {M.}~\bibnamefont {Sherwin}}, \ and\ \bibinfo {author}
  {\bibfnamefont {A.}~\bibnamefont {Small}},\ }\href {\doibase
  10.1103/PhysRevLett.83.4204} {\bibfield  {journal} {\bibinfo  {journal}
  {Phys. Rev. Lett.}\ }\textbf {\bibinfo {volume} {83}},\ \bibinfo {pages}
  {4204} (\bibinfo {year} {1999})}\BibitemShut {NoStop}%
\bibitem [{\citenamefont {Amasha}\ \emph {et~al.}(2008)\citenamefont {Amasha},
  \citenamefont {MacLean}, \citenamefont {Radu}, \citenamefont {Zumbühl},
  \citenamefont {Kastner}, \citenamefont {Hanson},\ and\ \citenamefont
  {Gossard}}]{amasha_electrical_2008}%
  \BibitemOpen
  \bibfield  {author} {\bibinfo {author} {\bibfnamefont {S.}~\bibnamefont
  {Amasha}}, \bibinfo {author} {\bibfnamefont {K.}~\bibnamefont {MacLean}},
  \bibinfo {author} {\bibfnamefont {I.~P.}\ \bibnamefont {Radu}}, \bibinfo
  {author} {\bibfnamefont {D.~M.}\ \bibnamefont {Zumbühl}}, \bibinfo {author}
  {\bibfnamefont {M.~A.}\ \bibnamefont {Kastner}}, \bibinfo {author}
  {\bibfnamefont {M.~P.}\ \bibnamefont {Hanson}}, \ and\ \bibinfo {author}
  {\bibfnamefont {A.~C.}\ \bibnamefont {Gossard}},\ }\href {\doibase
  10.1103/PhysRevLett.100.046803} {\bibfield  {journal} {\bibinfo  {journal}
  {Phys. Rev. Lett.}\ }\textbf {\bibinfo {volume} {100}},\ \bibinfo {pages}
  {046803} (\bibinfo {year} {2008})}\BibitemShut {NoStop}%
\bibitem [{\citenamefont {Scarlino}\ \emph {et~al.}(2014)\citenamefont
  {Scarlino}, \citenamefont {Kawakami}, \citenamefont {Stano}, \citenamefont
  {Shafiei}, \citenamefont {Reichl}, \citenamefont {Wegscheider},\ and\
  \citenamefont {Vandersypen}}]{scarlino_spin-relaxation_2014}%
  \BibitemOpen
  \bibfield  {author} {\bibinfo {author} {\bibfnamefont {P.}~\bibnamefont
  {Scarlino}}, \bibinfo {author} {\bibfnamefont {E.}~\bibnamefont {Kawakami}},
  \bibinfo {author} {\bibfnamefont {P.}~\bibnamefont {Stano}}, \bibinfo
  {author} {\bibfnamefont {M.}~\bibnamefont {Shafiei}}, \bibinfo {author}
  {\bibfnamefont {C.}~\bibnamefont {Reichl}}, \bibinfo {author} {\bibfnamefont
  {W.}~\bibnamefont {Wegscheider}}, \ and\ \bibinfo {author} {\bibfnamefont
  {L.}~\bibnamefont {Vandersypen}},\ }\href {\doibase
  10.1103/PhysRevLett.113.256802} {\bibfield  {journal} {\bibinfo  {journal}
  {Phys. Rev. Lett.}\ }\textbf {\bibinfo {volume} {113}},\ \bibinfo {pages}
  {256802} (\bibinfo {year} {2014})}\BibitemShut {NoStop}%
\bibitem [{\citenamefont {Könemann}\ \emph {et~al.}(2005)\citenamefont
  {Könemann}, \citenamefont {Haug}, \citenamefont {Maude}, \citenamefont
  {Fal’ko},\ and\ \citenamefont {Altshuler}}]{konemann_spin-orbit_2005}%
  \BibitemOpen
  \bibfield  {author} {\bibinfo {author} {\bibfnamefont {J.}~\bibnamefont
  {Könemann}}, \bibinfo {author} {\bibfnamefont {R.~J.}\ \bibnamefont {Haug}},
  \bibinfo {author} {\bibfnamefont {D.~K.}\ \bibnamefont {Maude}}, \bibinfo
  {author} {\bibfnamefont {V.~I.}\ \bibnamefont {Fal’ko}}, \ and\ \bibinfo
  {author} {\bibfnamefont {B.~L.}\ \bibnamefont {Altshuler}},\ }\href {\doibase
  10.1103/PhysRevLett.94.226404} {\bibfield  {journal} {\bibinfo  {journal}
  {Phys. Rev. Lett.}\ }\textbf {\bibinfo {volume} {94}},\ \bibinfo {pages}
  {226404} (\bibinfo {year} {2005})}\BibitemShut {NoStop}%
\bibitem [{\citenamefont {Golovach}\ \emph {et~al.}(2004)\citenamefont
  {Golovach}, \citenamefont {Khaetskii},\ and\ \citenamefont
  {Loss}}]{golovach_phonon-induced_2004}%
  \BibitemOpen
  \bibfield  {author} {\bibinfo {author} {\bibfnamefont {V.~N.}\ \bibnamefont
  {Golovach}}, \bibinfo {author} {\bibfnamefont {A.}~\bibnamefont {Khaetskii}},
  \ and\ \bibinfo {author} {\bibfnamefont {D.}~\bibnamefont {Loss}},\ }\href
  {\doibase 10.1103/PhysRevLett.93.016601} {\bibfield  {journal} {\bibinfo
  {journal} {Phys. Rev. Lett.}\ }\textbf {\bibinfo {volume} {93}},\ \bibinfo
  {pages} {016601} (\bibinfo {year} {2004})}\BibitemShut {NoStop}%
\bibitem [{\citenamefont {Golovach}\ \emph {et~al.}(2008)\citenamefont
  {Golovach}, \citenamefont {Khaetskii},\ and\ \citenamefont
  {Loss}}]{golovach_spin_2008}%
  \BibitemOpen
  \bibfield  {author} {\bibinfo {author} {\bibfnamefont {V.~N.}\ \bibnamefont
  {Golovach}}, \bibinfo {author} {\bibfnamefont {A.}~\bibnamefont {Khaetskii}},
  \ and\ \bibinfo {author} {\bibfnamefont {D.}~\bibnamefont {Loss}},\ }\href
  {\doibase 10.1103/PhysRevB.77.045328} {\bibfield  {journal} {\bibinfo
  {journal} {Phys. Rev. B}\ }\textbf {\bibinfo {volume} {77}},\ \bibinfo
  {pages} {045328} (\bibinfo {year} {2008})}\BibitemShut {NoStop}%
\bibitem [{\citenamefont {Khaetskii}\ and\ \citenamefont
  {Nazarov}(2000)}]{khaetskii_spin_2000}%
  \BibitemOpen
  \bibfield  {author} {\bibinfo {author} {\bibfnamefont {A.~V.}\ \bibnamefont
  {Khaetskii}}\ and\ \bibinfo {author} {\bibfnamefont {Y.~V.}\ \bibnamefont
  {Nazarov}},\ }\href {\doibase 10.1103/PhysRevB.61.12639} {\bibfield
  {journal} {\bibinfo  {journal} {Phys. Rev. B}\ }\textbf {\bibinfo {volume}
  {61}},\ \bibinfo {pages} {12639} (\bibinfo {year} {2000})}\BibitemShut
  {NoStop}%
\bibitem [{\citenamefont {Khaetskii}\ and\ \citenamefont
  {Nazarov}(2001)}]{khaetskii_spin-flip_2001}%
  \BibitemOpen
  \bibfield  {author} {\bibinfo {author} {\bibfnamefont {A.~V.}\ \bibnamefont
  {Khaetskii}}\ and\ \bibinfo {author} {\bibfnamefont {Y.~V.}\ \bibnamefont
  {Nazarov}},\ }\href {\doibase 10.1103/PhysRevB.64.125316} {\bibfield
  {journal} {\bibinfo  {journal} {Phys. Rev. B}\ }\textbf {\bibinfo {volume}
  {64}},\ \bibinfo {pages} {125316} (\bibinfo {year} {2001})}\BibitemShut
  {NoStop}%
\bibitem [{\citenamefont {Fal’ko}\ \emph {et~al.}(2005)\citenamefont
  {Fal’ko}, \citenamefont {Altshuler},\ and\ \citenamefont
  {Tsyplyatyev}}]{falko_anisotropy_2005}%
  \BibitemOpen
  \bibfield  {author} {\bibinfo {author} {\bibfnamefont {V.~I.}\ \bibnamefont
  {Fal’ko}}, \bibinfo {author} {\bibfnamefont {B.~L.}\ \bibnamefont
  {Altshuler}}, \ and\ \bibinfo {author} {\bibfnamefont {O.}~\bibnamefont
  {Tsyplyatyev}},\ }\href {\doibase 10.1103/PhysRevLett.95.076603} {\bibfield
  {journal} {\bibinfo  {journal} {Phys. Rev. Lett.}\ }\textbf {\bibinfo
  {volume} {95}},\ \bibinfo {pages} {076603} (\bibinfo {year}
  {2005})}\BibitemShut {NoStop}%
\bibitem [{\citenamefont {Stano}\ and\ \citenamefont
  {Fabian}(2006)}]{stano_orbital_2006}%
  \BibitemOpen
  \bibfield  {author} {\bibinfo {author} {\bibfnamefont {P.}~\bibnamefont
  {Stano}}\ and\ \bibinfo {author} {\bibfnamefont {J.}~\bibnamefont {Fabian}},\
  }\href {\doibase 10.1103/PhysRevB.74.045320} {\bibfield  {journal} {\bibinfo
  {journal} {Phys. Rev. B}\ }\textbf {\bibinfo {volume} {74}},\ \bibinfo
  {pages} {045320} (\bibinfo {year} {2006})}\BibitemShut {NoStop}%
\bibitem [{\citenamefont {Burkard}\ and\ \citenamefont
  {Loss}(2002)}]{burkard_cancellation_2002}%
  \BibitemOpen
  \bibfield  {author} {\bibinfo {author} {\bibfnamefont {G.}~\bibnamefont
  {Burkard}}\ and\ \bibinfo {author} {\bibfnamefont {D.}~\bibnamefont {Loss}},\
  }\href {\doibase 10.1103/PhysRevLett.88.047903} {\bibfield  {journal}
  {\bibinfo  {journal} {Phys. Rev. Lett.}\ }\textbf {\bibinfo {volume} {88}},\
  \bibinfo {pages} {047903} (\bibinfo {year} {2002})}\BibitemShut {NoStop}%
\bibitem [{\citenamefont {Stepanenko}\ \emph {et~al.}(2003)\citenamefont
  {Stepanenko}, \citenamefont {Bonesteel}, \citenamefont {DiVincenzo},
  \citenamefont {Burkard},\ and\ \citenamefont
  {Loss}}]{stepanenko_spin-orbit_2003}%
  \BibitemOpen
  \bibfield  {author} {\bibinfo {author} {\bibfnamefont {D.}~\bibnamefont
  {Stepanenko}}, \bibinfo {author} {\bibfnamefont {N.~E.}\ \bibnamefont
  {Bonesteel}}, \bibinfo {author} {\bibfnamefont {D.~P.}\ \bibnamefont
  {DiVincenzo}}, \bibinfo {author} {\bibfnamefont {G.}~\bibnamefont {Burkard}},
  \ and\ \bibinfo {author} {\bibfnamefont {D.}~\bibnamefont {Loss}},\ }\href
  {\doibase 10.1103/PhysRevB.68.115306} {\bibfield  {journal} {\bibinfo
  {journal} {Phys. Rev. B}\ }\textbf {\bibinfo {volume} {68}},\ \bibinfo
  {pages} {115306} (\bibinfo {year} {2003})}\BibitemShut {NoStop}%
\bibitem [{\citenamefont {Stepanenko}\ \emph {et~al.}(2012)\citenamefont
  {Stepanenko}, \citenamefont {Rudner}, \citenamefont {Halperin},\ and\
  \citenamefont {Loss}}]{stepanenko_singlet-triplet_2012}%
  \BibitemOpen
  \bibfield  {author} {\bibinfo {author} {\bibfnamefont {D.}~\bibnamefont
  {Stepanenko}}, \bibinfo {author} {\bibfnamefont {M.}~\bibnamefont {Rudner}},
  \bibinfo {author} {\bibfnamefont {B.~I.}\ \bibnamefont {Halperin}}, \ and\
  \bibinfo {author} {\bibfnamefont {D.}~\bibnamefont {Loss}},\ }\href {\doibase
  10.1103/PhysRevB.85.075416} {\bibfield  {journal} {\bibinfo  {journal} {Phys.
  Rev. B}\ }\textbf {\bibinfo {volume} {85}},\ \bibinfo {pages} {075416}
  (\bibinfo {year} {2012})}\BibitemShut {NoStop}%
\bibitem [{\citenamefont {Kavokin}(2001)}]{kavokin_anisotropic_2001}%
  \BibitemOpen
  \bibfield  {author} {\bibinfo {author} {\bibfnamefont {K.~V.}\ \bibnamefont
  {Kavokin}},\ }\href {\doibase 10.1103/PhysRevB.64.075305} {\bibfield
  {journal} {\bibinfo  {journal} {Phys. Rev. B}\ }\textbf {\bibinfo {volume}
  {64}},\ \bibinfo {pages} {075305} (\bibinfo {year} {2001})}\BibitemShut
  {NoStop}%
\bibitem [{\citenamefont {Danon}(2013)}]{danon_spin-flip_2013}%
  \BibitemOpen
  \bibfield  {author} {\bibinfo {author} {\bibfnamefont {J.}~\bibnamefont
  {Danon}},\ }\href {\doibase 10.1103/PhysRevB.88.075306} {\bibfield  {journal}
  {\bibinfo  {journal} {Phys. Rev. B}\ }\textbf {\bibinfo {volume} {88}},\
  \bibinfo {pages} {075306} (\bibinfo {year} {2013})}\BibitemShut {NoStop}%
\bibitem [{\citenamefont {Nichol}\ \emph {et~al.}(2015)\citenamefont {Nichol},
  \citenamefont {Harvey}, \citenamefont {Shulman}, \citenamefont {Pal},
  \citenamefont {Umansky}, \citenamefont {Rashba}, \citenamefont {Halperin},\
  and\ \citenamefont {Yacoby}}]{nichol_quenching_2015}%
  \BibitemOpen
  \bibfield  {author} {\bibinfo {author} {\bibfnamefont {J.~M.}\ \bibnamefont
  {Nichol}}, \bibinfo {author} {\bibfnamefont {S.~P.}\ \bibnamefont {Harvey}},
  \bibinfo {author} {\bibfnamefont {M.~D.}\ \bibnamefont {Shulman}}, \bibinfo
  {author} {\bibfnamefont {A.}~\bibnamefont {Pal}}, \bibinfo {author}
  {\bibfnamefont {V.}~\bibnamefont {Umansky}}, \bibinfo {author} {\bibfnamefont
  {E.~I.}\ \bibnamefont {Rashba}}, \bibinfo {author} {\bibfnamefont {B.~I.}\
  \bibnamefont {Halperin}}, \ and\ \bibinfo {author} {\bibfnamefont
  {A.}~\bibnamefont {Yacoby}},\ }\href {\doibase 10.1038/ncomms8682} {\bibfield
   {journal} {\bibinfo  {journal} {Nature Communications}\ }\textbf {\bibinfo
  {volume} {6}},\ \bibinfo {pages} {7682} (\bibinfo {year} {2015})}\BibitemShut
  {NoStop}%
\bibitem [{\citenamefont {Stano}\ and\ \citenamefont
  {Fabian}(2005)}]{stano_spin-orbit_2005}%
  \BibitemOpen
  \bibfield  {author} {\bibinfo {author} {\bibfnamefont {P.}~\bibnamefont
  {Stano}}\ and\ \bibinfo {author} {\bibfnamefont {J.}~\bibnamefont {Fabian}},\
  }\href {\doibase 10.1103/PhysRevB.72.155410} {\bibfield  {journal} {\bibinfo
  {journal} {Phys. Rev. B}\ }\textbf {\bibinfo {volume} {72}},\ \bibinfo
  {pages} {155410} (\bibinfo {year} {2005})}\BibitemShut {NoStop}%
\bibitem [{\citenamefont {Hofmann}\ \emph {et~al.}(2016)\citenamefont
  {Hofmann}, \citenamefont {Maisi}, \citenamefont {Gold}, \citenamefont
  {Krähenmann}, \citenamefont {Rössler}, \citenamefont {Basset},
  \citenamefont {Märki}, \citenamefont {Reichl}, \citenamefont {Wegscheider},
  \citenamefont {Ensslin},\ and\ \citenamefont {Ihn}}]{hofmann_measuring_2016}%
  \BibitemOpen
  \bibfield  {author} {\bibinfo {author} {\bibfnamefont {A.}~\bibnamefont
  {Hofmann}}, \bibinfo {author} {\bibfnamefont {V.}~\bibnamefont {Maisi}},
  \bibinfo {author} {\bibfnamefont {C.}~\bibnamefont {Gold}}, \bibinfo {author}
  {\bibfnamefont {T.}~\bibnamefont {Krähenmann}}, \bibinfo {author}
  {\bibfnamefont {C.}~\bibnamefont {Rössler}}, \bibinfo {author}
  {\bibfnamefont {J.}~\bibnamefont {Basset}}, \bibinfo {author} {\bibfnamefont
  {P.}~\bibnamefont {Märki}}, \bibinfo {author} {\bibfnamefont
  {C.}~\bibnamefont {Reichl}}, \bibinfo {author} {\bibfnamefont
  {W.}~\bibnamefont {Wegscheider}}, \bibinfo {author} {\bibfnamefont
  {K.}~\bibnamefont {Ensslin}}, \ and\ \bibinfo {author} {\bibfnamefont
  {T.}~\bibnamefont {Ihn}},\ }\href {\doibase 10.1103/PhysRevLett.117.206803}
  {\bibfield  {journal} {\bibinfo  {journal} {Phys. Rev. Lett.}\ }\textbf
  {\bibinfo {volume} {117}},\ \bibinfo {pages} {206803} (\bibinfo {year}
  {2016})}\BibitemShut {NoStop}%
\bibitem [{\citenamefont {Hanson}\ \emph {et~al.}(2007)\citenamefont {Hanson},
  \citenamefont {Kouwenhoven}, \citenamefont {Petta}, \citenamefont {Tarucha},\
  and\ \citenamefont {Vandersypen}}]{hanson_spins_2007}%
  \BibitemOpen
  \bibfield  {author} {\bibinfo {author} {\bibfnamefont {R.}~\bibnamefont
  {Hanson}}, \bibinfo {author} {\bibfnamefont {L.~P.}\ \bibnamefont
  {Kouwenhoven}}, \bibinfo {author} {\bibfnamefont {J.~R.}\ \bibnamefont
  {Petta}}, \bibinfo {author} {\bibfnamefont {S.}~\bibnamefont {Tarucha}}, \
  and\ \bibinfo {author} {\bibfnamefont {L.~M.~K.}\ \bibnamefont
  {Vandersypen}},\ }\href {\doibase 10.1103/RevModPhys.79.1217} {\bibfield
  {journal} {\bibinfo  {journal} {Rev. Mod. Phys.}\ }\textbf {\bibinfo {volume}
  {79}},\ \bibinfo {pages} {1217} (\bibinfo {year} {2007})}\BibitemShut
  {NoStop}%
\bibitem [{\citenamefont {Reichl}(2016)}]{reichl_modern_2016}%
  \BibitemOpen
  \bibfield  {author} {\bibinfo {author} {\bibfnamefont {L.~E.}\ \bibnamefont
  {Reichl}},\ }\href@noop {} {\emph {\bibinfo {title} {A modern course in
  statistical physics}}},\ \bibinfo {edition} {4th}\ ed.\ (\bibinfo
  {publisher} {Wiley-VCH Verlag GmbH \& Co. KGaA},\ \bibinfo {address}
  {Weinheim},\ \bibinfo {year} {2016})\ \bibinfo {note} {oCLC:
  927845638}\BibitemShut {NoStop}%
\bibitem [{\citenamefont {Maisi}\ \emph {et~al.}(2016)\citenamefont {Maisi},
  \citenamefont {Hofmann}, \citenamefont {Röösli}, \citenamefont {Basset},
  \citenamefont {Reichl}, \citenamefont {Wegscheider}, \citenamefont {Ihn},\
  and\ \citenamefont {Ensslin}}]{maisi_spin-orbit_2016}%
  \BibitemOpen
  \bibfield  {author} {\bibinfo {author} {\bibfnamefont {V.~F.}\ \bibnamefont
  {Maisi}}, \bibinfo {author} {\bibfnamefont {A.}~\bibnamefont {Hofmann}},
  \bibinfo {author} {\bibfnamefont {M.}~\bibnamefont {Röösli}}, \bibinfo
  {author} {\bibfnamefont {J.}~\bibnamefont {Basset}}, \bibinfo {author}
  {\bibfnamefont {C.}~\bibnamefont {Reichl}}, \bibinfo {author} {\bibfnamefont
  {W.}~\bibnamefont {Wegscheider}}, \bibinfo {author} {\bibfnamefont
  {T.}~\bibnamefont {Ihn}}, \ and\ \bibinfo {author} {\bibfnamefont
  {K.}~\bibnamefont {Ensslin}},\ }\href {\doibase
  10.1103/PhysRevLett.116.136803} {\bibfield  {journal} {\bibinfo  {journal}
  {Physical Review Letters}\ }\textbf {\bibinfo {volume} {116}},\ \bibinfo
  {pages} {136803} (\bibinfo {year} {2016})}\BibitemShut {NoStop}%
\bibitem [{\citenamefont {Fujita}\ \emph {et~al.}(2016)\citenamefont {Fujita},
  \citenamefont {Stano}, \citenamefont {Allison}, \citenamefont {Morimoto},
  \citenamefont {Sato}, \citenamefont {Larsson}, \citenamefont {Park},
  \citenamefont {Ludwig}, \citenamefont {Wieck}, \citenamefont {Oiwa},\ and\
  \citenamefont {Tarucha}}]{fujita_signatures_2016}%
  \BibitemOpen
  \bibfield  {author} {\bibinfo {author} {\bibfnamefont {T.}~\bibnamefont
  {Fujita}}, \bibinfo {author} {\bibfnamefont {P.}~\bibnamefont {Stano}},
  \bibinfo {author} {\bibfnamefont {G.}~\bibnamefont {Allison}}, \bibinfo
  {author} {\bibfnamefont {K.}~\bibnamefont {Morimoto}}, \bibinfo {author}
  {\bibfnamefont {Y.}~\bibnamefont {Sato}}, \bibinfo {author} {\bibfnamefont
  {M.}~\bibnamefont {Larsson}}, \bibinfo {author} {\bibfnamefont {J.-H.}\
  \bibnamefont {Park}}, \bibinfo {author} {\bibfnamefont {A.}~\bibnamefont
  {Ludwig}}, \bibinfo {author} {\bibfnamefont {A.}~\bibnamefont {Wieck}},
  \bibinfo {author} {\bibfnamefont {A.}~\bibnamefont {Oiwa}}, \ and\ \bibinfo
  {author} {\bibfnamefont {S.}~\bibnamefont {Tarucha}},\ }\href {\doibase
  10.1103/PhysRevLett.117.206802} {\bibfield  {journal} {\bibinfo  {journal}
  {Phys. Rev. Lett.}\ }\textbf {\bibinfo {volume} {117}},\ \bibinfo {pages}
  {206802} (\bibinfo {year} {2016})}\BibitemShut {NoStop}%
\bibitem [{\citenamefont {Studer}\ \emph {et~al.}(2009)\citenamefont {Studer},
  \citenamefont {Salis}, \citenamefont {Ensslin}, \citenamefont {Driscoll},\
  and\ \citenamefont {Gossard}}]{studer_gate-controlled_2009}%
  \BibitemOpen
  \bibfield  {author} {\bibinfo {author} {\bibfnamefont {M.}~\bibnamefont
  {Studer}}, \bibinfo {author} {\bibfnamefont {G.}~\bibnamefont {Salis}},
  \bibinfo {author} {\bibfnamefont {K.}~\bibnamefont {Ensslin}}, \bibinfo
  {author} {\bibfnamefont {D.~C.}\ \bibnamefont {Driscoll}}, \ and\ \bibinfo
  {author} {\bibfnamefont {A.~C.}\ \bibnamefont {Gossard}},\ }\href {\doibase
  10.1103/PhysRevLett.103.027201} {\bibfield  {journal} {\bibinfo  {journal}
  {Phys. Rev. Lett.}\ }\textbf {\bibinfo {volume} {103}},\ \bibinfo {pages}
  {027201} (\bibinfo {year} {2009})}\BibitemShut {NoStop}%
\bibitem [{\citenamefont {Dettwiler}\ \emph {et~al.}(2017)\citenamefont
  {Dettwiler}, \citenamefont {Fu}, \citenamefont {Mack}, \citenamefont
  {Weigele}, \citenamefont {Egues}, \citenamefont {Awschalom},\ and\
  \citenamefont {Zumbühl}}]{dettwiler_stretchable_2017}%
  \BibitemOpen
  \bibfield  {author} {\bibinfo {author} {\bibfnamefont {F.}~\bibnamefont
  {Dettwiler}}, \bibinfo {author} {\bibfnamefont {J.}~\bibnamefont {Fu}},
  \bibinfo {author} {\bibfnamefont {S.}~\bibnamefont {Mack}}, \bibinfo {author}
  {\bibfnamefont {P.~J.}\ \bibnamefont {Weigele}}, \bibinfo {author}
  {\bibfnamefont {J.~C.}\ \bibnamefont {Egues}}, \bibinfo {author}
  {\bibfnamefont {D.~D.}\ \bibnamefont {Awschalom}}, \ and\ \bibinfo {author}
  {\bibfnamefont {D.~M.}\ \bibnamefont {Zumbühl}},\ }\href
  {http://arxiv.org/abs/1702.05190} {\bibfield  {journal} {\bibinfo  {journal}
  {arXiv:1702.05190 [cond-mat]}\ } (\bibinfo {year} {2017})},\ \bibinfo {note}
  {arXiv: 1702.05190}\BibitemShut {NoStop}%
\bibitem [{\citenamefont {Schliemann}\ \emph {et~al.}(2003)\citenamefont
  {Schliemann}, \citenamefont {Egues},\ and\ \citenamefont
  {Loss}}]{schliemann_nonballistic_2003}%
  \BibitemOpen
  \bibfield  {author} {\bibinfo {author} {\bibfnamefont {J.}~\bibnamefont
  {Schliemann}}, \bibinfo {author} {\bibfnamefont {J.~C.}\ \bibnamefont
  {Egues}}, \ and\ \bibinfo {author} {\bibfnamefont {D.}~\bibnamefont {Loss}},\
  }\href {\doibase 10.1103/PhysRevLett.90.146801} {\bibfield  {journal}
  {\bibinfo  {journal} {Phys. Rev. Lett.}\ }\textbf {\bibinfo {volume} {90}},\
  \bibinfo {pages} {146801} (\bibinfo {year} {2003})}\BibitemShut {NoStop}%
\bibitem [{\citenamefont {Vandersypen}\ \emph {et~al.}(2004)\citenamefont
  {Vandersypen}, \citenamefont {Elzerman}, \citenamefont {Schouten},
  \citenamefont {Beveren}, \citenamefont {Hanson},\ and\ \citenamefont
  {Kouwenhoven}}]{vandersypen_real-time_2004}%
  \BibitemOpen
  \bibfield  {author} {\bibinfo {author} {\bibfnamefont {L.~M.~K.}\
  \bibnamefont {Vandersypen}}, \bibinfo {author} {\bibfnamefont {J.~M.}\
  \bibnamefont {Elzerman}}, \bibinfo {author} {\bibfnamefont {R.~N.}\
  \bibnamefont {Schouten}}, \bibinfo {author} {\bibfnamefont {L.~H. W.~v.}\
  \bibnamefont {Beveren}}, \bibinfo {author} {\bibfnamefont {R.}~\bibnamefont
  {Hanson}}, \ and\ \bibinfo {author} {\bibfnamefont {L.~P.}\ \bibnamefont
  {Kouwenhoven}},\ }\href {\doibase 10.1063/1.1815041} {\bibfield  {journal}
  {\bibinfo  {journal} {Applied Physics Letters}\ }\textbf {\bibinfo {volume}
  {85}},\ \bibinfo {pages} {4394} (\bibinfo {year} {2004})}\BibitemShut
  {NoStop}%
\bibitem [{\citenamefont {Schleser}\ \emph {et~al.}(2004)\citenamefont
  {Schleser}, \citenamefont {Ruh}, \citenamefont {Ihn}, \citenamefont
  {Ensslin}, \citenamefont {Driscoll},\ and\ \citenamefont
  {Gossard}}]{schleser_time-resolved_2004}%
  \BibitemOpen
  \bibfield  {author} {\bibinfo {author} {\bibfnamefont {R.}~\bibnamefont
  {Schleser}}, \bibinfo {author} {\bibfnamefont {E.}~\bibnamefont {Ruh}},
  \bibinfo {author} {\bibfnamefont {T.}~\bibnamefont {Ihn}}, \bibinfo {author}
  {\bibfnamefont {K.}~\bibnamefont {Ensslin}}, \bibinfo {author} {\bibfnamefont
  {D.~C.}\ \bibnamefont {Driscoll}}, \ and\ \bibinfo {author} {\bibfnamefont
  {A.~C.}\ \bibnamefont {Gossard}},\ }\href {\doibase 10.1063/1.1784875}
  {\bibfield  {journal} {\bibinfo  {journal} {Applied Physics Letters}\
  }\textbf {\bibinfo {volume} {85}},\ \bibinfo {pages} {2005} (\bibinfo {year}
  {2004})}\BibitemShut {NoStop}%
\bibitem [{\citenamefont {Jouravlev}\ and\ \citenamefont
  {Nazarov}(2006)}]{jouravlev_electron_2006}%
  \BibitemOpen
  \bibfield  {author} {\bibinfo {author} {\bibfnamefont {O.~N.}\ \bibnamefont
  {Jouravlev}}\ and\ \bibinfo {author} {\bibfnamefont {Y.~V.}\ \bibnamefont
  {Nazarov}},\ }\href {\doibase 10.1103/PhysRevLett.96.176804} {\bibfield
  {journal} {\bibinfo  {journal} {Phys. Rev. Lett.}\ }\textbf {\bibinfo
  {volume} {96}},\ \bibinfo {pages} {176804} (\bibinfo {year}
  {2006})}\BibitemShut {NoStop}%
\bibitem [{\citenamefont {Koppens}\ \emph {et~al.}(2005)\citenamefont
  {Koppens}, \citenamefont {Folk}, \citenamefont {Elzerman}, \citenamefont
  {Hanson}, \citenamefont {Beveren}, \citenamefont {Vink}, \citenamefont
  {Tranitz}, \citenamefont {Wegscheider}, \citenamefont {Kouwenhoven},\ and\
  \citenamefont {Vandersypen}}]{koppens_control_2005}%
  \BibitemOpen
  \bibfield  {author} {\bibinfo {author} {\bibfnamefont {F.~H.~L.}\
  \bibnamefont {Koppens}}, \bibinfo {author} {\bibfnamefont {J.~A.}\
  \bibnamefont {Folk}}, \bibinfo {author} {\bibfnamefont {J.~M.}\ \bibnamefont
  {Elzerman}}, \bibinfo {author} {\bibfnamefont {R.}~\bibnamefont {Hanson}},
  \bibinfo {author} {\bibfnamefont {L.~H. W.~v.}\ \bibnamefont {Beveren}},
  \bibinfo {author} {\bibfnamefont {I.~T.}\ \bibnamefont {Vink}}, \bibinfo
  {author} {\bibfnamefont {H.~P.}\ \bibnamefont {Tranitz}}, \bibinfo {author}
  {\bibfnamefont {W.}~\bibnamefont {Wegscheider}}, \bibinfo {author}
  {\bibfnamefont {L.~P.}\ \bibnamefont {Kouwenhoven}}, \ and\ \bibinfo {author}
  {\bibfnamefont {L.~M.~K.}\ \bibnamefont {Vandersypen}},\ }\href {\doibase
  10.1126/science.1113719} {\bibfield  {journal} {\bibinfo  {journal}
  {Science}\ }\textbf {\bibinfo {volume} {309}},\ \bibinfo {pages} {1346}
  (\bibinfo {year} {2005})}\BibitemShut {NoStop}%
\bibitem [{\citenamefont {Ono}\ \emph {et~al.}(2002)\citenamefont {Ono},
  \citenamefont {Austing}, \citenamefont {Tokura},\ and\ \citenamefont
  {Tarucha}}]{ono_current_2002}%
  \BibitemOpen
  \bibfield  {author} {\bibinfo {author} {\bibfnamefont {K.}~\bibnamefont
  {Ono}}, \bibinfo {author} {\bibfnamefont {D.~G.}\ \bibnamefont {Austing}},
  \bibinfo {author} {\bibfnamefont {Y.}~\bibnamefont {Tokura}}, \ and\ \bibinfo
  {author} {\bibfnamefont {S.}~\bibnamefont {Tarucha}},\ }\href {\doibase
  10.1126/science.1070958} {\bibfield  {journal} {\bibinfo  {journal}
  {Science}\ }\textbf {\bibinfo {volume} {297}},\ \bibinfo {pages} {1313}
  (\bibinfo {year} {2002})}\BibitemShut {NoStop}%
\bibitem [{\citenamefont {Johnson}\ \emph {et~al.}(2005)\citenamefont
  {Johnson}, \citenamefont {Petta}, \citenamefont {Marcus}, \citenamefont
  {Hanson},\ and\ \citenamefont {Gossard}}]{johnson_singlet-triplet_2005}%
  \BibitemOpen
  \bibfield  {author} {\bibinfo {author} {\bibfnamefont {A.~C.}\ \bibnamefont
  {Johnson}}, \bibinfo {author} {\bibfnamefont {J.~R.}\ \bibnamefont {Petta}},
  \bibinfo {author} {\bibfnamefont {C.~M.}\ \bibnamefont {Marcus}}, \bibinfo
  {author} {\bibfnamefont {M.~P.}\ \bibnamefont {Hanson}}, \ and\ \bibinfo
  {author} {\bibfnamefont {A.~C.}\ \bibnamefont {Gossard}},\ }\href {\doibase
  10.1103/PhysRevB.72.165308} {\bibfield  {journal} {\bibinfo  {journal} {Phys.
  Rev. B}\ }\textbf {\bibinfo {volume} {72}},\ \bibinfo {pages} {165308}
  (\bibinfo {year} {2005})}\BibitemShut {NoStop}%
\bibitem [{\citenamefont {Fujita}\ \emph {et~al.}(2015)\citenamefont {Fujita},
  \citenamefont {Morimoto}, \citenamefont {Kiyama}, \citenamefont {Allison},
  \citenamefont {Larsson}, \citenamefont {Ludwig}, \citenamefont {Valentin},
  \citenamefont {Wieck}, \citenamefont {Oiwa},\ and\ \citenamefont
  {Tarucha}}]{fujita_single_2015}%
  \BibitemOpen
  \bibfield  {author} {\bibinfo {author} {\bibfnamefont {T.}~\bibnamefont
  {Fujita}}, \bibinfo {author} {\bibfnamefont {K.}~\bibnamefont {Morimoto}},
  \bibinfo {author} {\bibfnamefont {H.}~\bibnamefont {Kiyama}}, \bibinfo
  {author} {\bibfnamefont {G.}~\bibnamefont {Allison}}, \bibinfo {author}
  {\bibfnamefont {M.}~\bibnamefont {Larsson}}, \bibinfo {author} {\bibfnamefont
  {A.}~\bibnamefont {Ludwig}}, \bibinfo {author} {\bibfnamefont {S.~R.}\
  \bibnamefont {Valentin}}, \bibinfo {author} {\bibfnamefont {A.~D.}\
  \bibnamefont {Wieck}}, \bibinfo {author} {\bibfnamefont {A.}~\bibnamefont
  {Oiwa}}, \ and\ \bibinfo {author} {\bibfnamefont {S.}~\bibnamefont
  {Tarucha}},\ }\href {http://arxiv.org/abs/1504.03696} {\bibfield  {journal}
  {\bibinfo  {journal} {arXiv:1504.03696 [cond-mat, physics:quant-ph]}\ }
  (\bibinfo {year} {2015})},\ \bibinfo {note} {arXiv: 1504.03696}\BibitemShut
  {NoStop}%
\end{thebibliography}%
\bibliographystyle{apsrev4-1}
\end{document}